**Evolution and Transformation of Scientific Knowledge over the *Sphaera* Corpus: A Network Study**

Maryam Zamani,[1] Alejandro Tejedor,[2] Malte Vogl,[3] Florian Kräutli,[3] Matteo Valleriani,[3, 4, 5] and Holger Kantz[1]

[1] Max Planck Institute for the Physics of Complex Systems, Dresden, Germany
[2] Department of Science and Engineering, Sorbonne University Abu Dhabi, Abu Dhabi, United Arab Emirates
[3] Max Planck Institute for the History of Science, Berlin, Germany
[4] Technische Universität Berlin, Germany
[5] Tel Aviv University, Israel

**Abstract**

We investigated the evolution and transformation of scientific knowledge in the early modern period, analyzing more than 350 different editions of textbooks used for teaching astronomy in European universities from the late fifteenth century to mid-seventeenth century. These historical sources constitute the *Sphaera Corpus*. By examining different semantic relations among individual parts of each edition on record, we built a multiplex network consisting of six layers, as well as the aggregated network built from the superposition of all the layers. The network analysis reveals the emergence of five different communities. The contribution of each layer in shaping the communities and the properties of each community are studied. The most influential books in the corpus are found by calculating the average age of all the out-going and in-coming links for each book. A small group of editions is identified as a transmitter of knowledge as they bridge past knowledge to the future through a long temporal interval. Our analysis, moreover, identifies the most disruptive books. These books introduce new knowledge that is then adopted by almost all the books published afterwards until the end of the whole period of study. The historical research on the content of the identified books, as an empirical test, finally corroborates the results of all our analyses.

---

**Introduction**

How does scientific knowledge evolve [1-5]? How did it evolve in the past, for instance, during a specific period of time? How does knowledge transform? Which epistemic mechanisms make knowledge homogenous and standardized and which causes it to differentiate and diverge [6-8]? And finally, how are these transformations intertwined with the process of transmission and preservation of knowledge over time [9-11]? These relevant and fascinating questions can be approached by means of network study as it provides efficient and appropriate investigative tools [12-14]. By means of this approach, scholars have, for instance, studied how knowledge has been transferred among different disciplines or shared among scholars. Knowledge is considered to develop through scientific discussions and networks of interaction between books and papers [15,16], and researchers. For this type of study, therefore, citation networks have primarily been used [17,18].

The aim of this paper is to investigate the process of development of scientific knowledge in the early modern period, a time in which scholars did not cite each other in the same way as in the current epoch. The source of this study is a collection of 356 university textbooks [19], which is called *Sphaera* corpus. These textbooks were used in introductory astronomy and cosmology classes in liberal arts faculties across Europe during the early modern period. The treatises have been collected on the basis of one historical condition, namely, that they contain a specific text compiled during the thirteenth century by Johannes de Sacrobosco, the *Tractatus de sphaera*. This text was written in Paris in the frame of the newly founded university, and it was used for a qualitative, non-mathematical introduction to geocentric cosmology. The treatise was new and original in its structure and did not have the typical design of the medieval commentary. Likely for these reasons, it became a mandatory textbook in all European universities and remained as such until the 17$_{th}$ century [20]. The network of interaction between these books is built in six different layers based on their contents. In this approach, each book is represented as a node, and two books are connected if they share the same parts of texts according to a sophisticated ontology described in the next section.

In a recent publication, we focused on the same corpus. We considered five layers of interactions between these books and the influence of each book in the corpus was studied using normalized out-degree of each book in the corresponding aggregated graph (with all the layers collapsed into one), and the results are discussed mainly from the historical point of view [21]. While some of these results will be briefly reviewed, we structured the data differently in the present work by adding one more layer in order to better differentiate historical sources. After restructuring the network, we restarted here with the same out- and in-degree method of analysis. We identified the emergence of five different communities of books (epistemic communities) where each community corresponds to a branch in the normalized out-degree graph. Communities are recognized by the Louvain community detection algorithm [22]. The density of links in each community is much higher than the links between communities which means that books in each community are highly related to one another. Finally, we measure the average age of all the links approaching and stemming from each book, and the results are then compared with the same measurement as the fully connected graph. This comparison detects several significant books that have the role of passing knowledge from past periods to future publications. This analysis also reveals the most disruptive books that are not influenced from the past (the average age of incoming links is very low) but have a very high impact on later publications (affecting almost all subsequently published books) collected in the corpus.

**1. Data and Network**

As mentioned, we considered a corpus of 356 historical sources [23]. To keep consistency in the dataset [24], only the printed versions of the textbooks are considered while the manuscripts are ignored as no census of them has ever been realized. Sacrobosco's treatise was printed in 1472 for the first time. While this first printed version contained only the text of Sacrobosco, soon thereafter it was compiled with a variety of other texts that commented on, criticized, or enriched the original text either in part or in entirety. This process continued until the seventeenth century when the relevance of the original medieval tract in the university classes of Europe decreased. The last treatise collected in the corpus analyzed here was printed in 1650, meaning the corpus covers a period of 178 years and as such is the time span of our study. Finally, though with very different rates of production, the textbooks were printed in 40 European cities [25].

We collected all relevant pieces of information concerning all the texts that were printed together with the original treatise of Sacrobosco. In this way we "atomized" all the 356 editions into what we called *text parts*. Sacrobosco's treatise, for instance, is a text part in and of itself while other text parts may be new, original scientific treatises, commentaries, translations, fragments, or paratexts of more literary nature [21].

We developed a taxonomy of text parts that is comprised of the following categories: *original text*, *commentary*, and *translation*. The last two are referred to under the term *adaption (part)* as they relate to a text part. While an original text can be both a new text and a text that is being printed for the first time, a commentary is a text that expands on specific passages of an original text in positive or negative way. This method of producing scientific texts was the standard way to partake in scientific debates in terms of circulating publications. In the print of a commentary, the original text (in this case, also called *text of reference*) was printed alongside the commentary, eventually in a different font and font size and including a scientific illustration. The illustration itself also carried scientific knowledge, though perhaps not completely overlapping with the knowledge expressed in the text and can therefore to be interpreted as a form of commentary as well (Fig. 1).

Because a subgroup of these text parts tends to reoccur in later editions, it was possible to build networks based on semantic relations between books in the corpus based on the fact that they were sharing the same content, at least partially. To correctly understand the semantic structures, we made use of while building the networks, however, it is first necessary to briefly recall the taxonomy according to which the editions and all the works (and not only their text parts) are organized. An extended exposition of the taxonomy can be found in [25]. The corpus consists of five different sorts of textbook editions. The first is constituted of printed books that contain only the original treatise of Sacrobosco, the second of works that contain the original tract and a commentary (as shown in Fig. 1), the third of textbooks that contain the original texts and other texts that function as commentaries in the entire book but are added one after the other and not printed on the same page, and the fourth by a combination of the second and the third typology. Finally, the fifth is consists of editions that we have classified as *adaptions (book)*, and is the most relevant in this argument. This category denotes a relevant subgroup in terms of size because it includes 124 different editions. These books constitute a peculiar subgroup of editions within the corpus. They contain at least one text part that is considered to be *strongly influenced* by a reference text. In most cases, this is the same *Tractatus de sphaera* of Sacrobosco, but they significantly do not contain the treatise and do not comment directly on it. They are considered to be influenced by it because they discuss either all or a majority of the same subjects. Moreover, they discuss them by following either the same or a very similar order. Finally, they largely make use of the same visual apparatus [26]. Nevertheless, they often contain different scientific arguments and different views, though on the same subjects. These books therefore represent the first strong departure from the tradition of textbooks associated with Sacrobosco's treatise. Yet, they still belong to this tradition because they do not change its identity and do not argue in favor of a structurally different science of cosmology. They are, in other terms, at the boundaries of the corpus from a content-related point of view.

When the corpus was analyzed in [21] not all possible semantic relations were exploited, including those relations that concerned editions grouped under the label adaptions (book). They were identified but then ignored. The present work instead considers all possible semantic relations among text parts and correspondingly increases the number of layers of the network.

Based on both the taxonomy of the text parts and of the entire editions, a multiplex network with six different layers is therefore produced on the basis of the following semantic structure (Fig. 2):

**Layer se13**- *Same Original Part*: Two books are in relation to each other if they contain exactly the same original part, for instance the same dedication letter or the same treatise in the same language and by the same author.

**Layer se14**- *Same Adaption (part)*: Two books are in relation to each other if they contain exactly the same part and this text part is an adaption, for instance a commentary on the *Tractatus* of Sacrobosco. For instance, the two books hdl.handle.net/21.11103/sphaera.101112 and hdl.handle.net/21.11103/sphaera.101056 are in relation to each other because they both contain Élie Vinet's commentary on the *Sphere* of Sacrobosco.

**Layer se15**- *Translated Same Original Part*: Two books are related to each other when they both contain a translation of the same original part. The translations do not have to be into the same language.

**Layer se16**- *Annotated Same Original Part*: Two books are related to each other if they both contain commentaries that are not the same but are on the same original part, as for instance the commentary of Francesco Capuano, published in the source book hdl.handle.net/21.11103/sphaera.100047 and the commentary of Francesco Giuntini, published in the target book hdl.handle.net/21.11103/sphaera.101101; both commentaries address the same original part, namely the *Theorica novae planetarum* by Peuerbach.

**Layer se17**- *Annotated Same Adaption*: Two books are related to each other if they both contain commentaries that are not the same but are on the same adaption, which is in turn a commentary on or a translation of an original text part. For instance, the source book hdl.handle.net/21.11103/sphaera.101114 is related to the target book

hdl.handle.net/21.11103/sphaera.100656 because they respectively contain Francesco Giuntini's commentary and Alberto Hero's commentary on Élie Vinet's adaption of Sacrobosco's S*phaera*, the latter an original part. If two parts that are the same adaption (part) are present in two editions, then the corresponding relation between the two editions is listed in layer se14, as it falls under the category same adaption.

**Layer se19**- *Adaption (Book) Influenced by Same Original Part*: Two books are related to each other if, first, these two books are adaptions in the sense described above and second, if they contain a text part which is *strongly influenced* by the same original text part. For instance, the source book hdl.handle.net/21.11103/sphaera.100190 is related to the target book hdl.handle.net/21.11103/sphaera.101303 because they are both adaptions (book).

Furthermore, the first contains Kaspar Peucer's four chapters on *elementorum sphaeriricorum* as a text part, and the second includes *De Sphaera et primis Astronomiae rudimentis libellus* by Cornelius Valerius also as a text part, and these two text parts are adaptions (strongly influenced by) the *Tractatus de sphaera* of Johannes de Sacrobosco.

The aggregated graph is built from the superposition of all the layers, and therefore contains all the editions of the corpus and all the semantic relations between them as described above. Table 1 shows the information concerned with the number of nodes and edges in the different layers of the multiplex network and in the aggregated graph. Edges in all layers and their superposition graph are directed, and directionality is based on chronological ordering from the older edition (source) to the newer one (target).

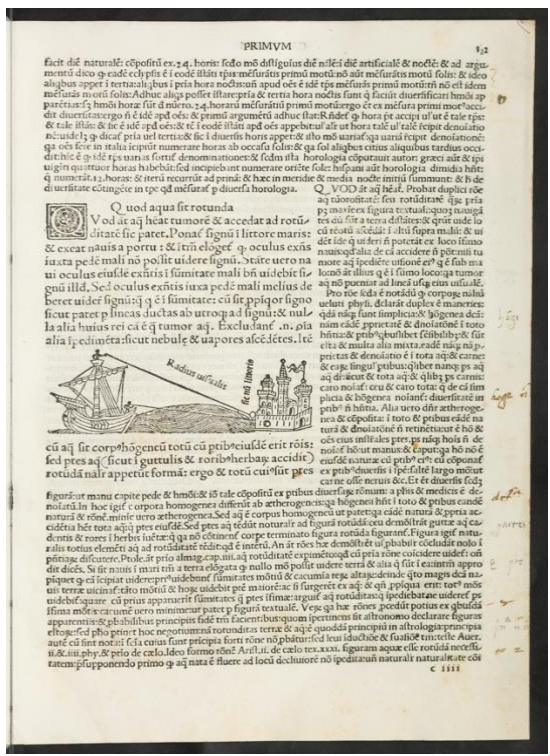

**Fig. 1.** Demonstration of the spherical shape of the water around the element earth. From Sacrobosco, Johannes de, Pierre d'Ailly, Francesco Capuano, Robert Grosseteste, Jacques Lefèvre d'Etaples, Joannes Regiomontanus, Georg von Peurbach, and Bartolomeo Vespucci. *Nota eorum quæ in hoc libro continentur... Textvs Sphaerae Ioannis De Sacro Bvsto.* Venice: Joannes Ruberus and Bernardinus Vercellensis for Giunta, 1508, p. 12r.

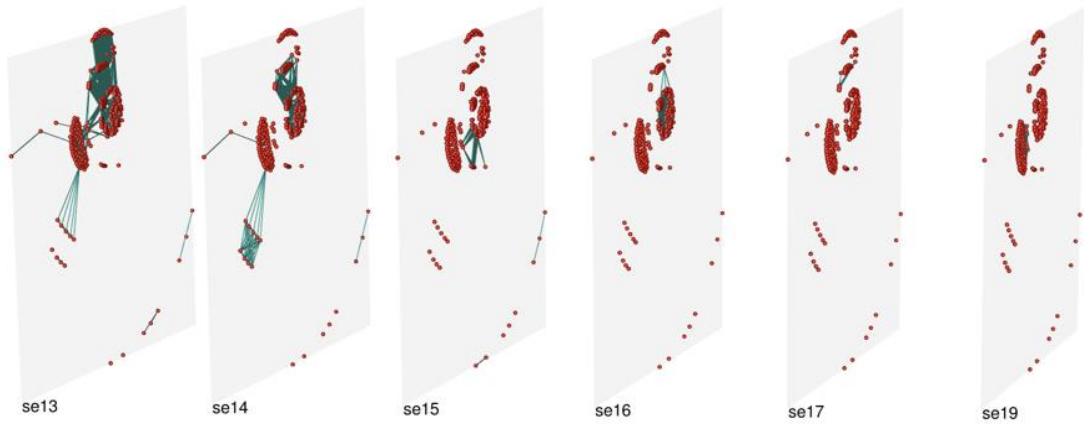

**Figure 2.** Multiplex network of the editions of the *Sphaera* corpus organized in six different layers. In each layer, books are connected on the basis of a specific semantic relation. Visualization realized by means of muxViz (muxViz.net) [27].

|            | Number of Nodes | Number of Edges |
|------------|-----------------|-----------------|
| se13       | 321             | 5681            |
| se14       | 199             | 2173            |
| se15       | 43              | 341             |
| se16       | 183             | 15326           |
| se17       | 8               | 15              |
| se19       | 85              | 3312            |
| Aggregated | 356             | 23586           |

**Table 1.** Number of editions (nodes) and edges (semantic relations) in each layer of the multiplex network and the corresponding aggregated graph.

## 2. Analyzing Data

In this section we describe the methods for analyzing the dataset. First, we describe the normalized out-degree and in-degree analysis in the aggregated graph to measure the influence of each edition on the other editions. As described in the last section, the aggregated graph is built by making a superposition of all the layers. The connections between books (nodes) are based on their semantic relations; there is a directed link (based on chronological ordering) between two books if they share at least one semantic part. Out-degree is the number of out-going links from each node and in-degree is the total number of incoming links to each node. We observe five different communities in the aggregated graph and study the properties of each community in subsection 2.2. Finally, we investigate the depth of influence that each book gets from the past by measuring the average age of the incoming links and we study their corresponding impact on future editions by measuring the average age of out-going links.

## 2.1 Normalized out-degree and in-degree

We study the direct influence of each book in the corpus by computing the normalized out-degree, which is the ratio of the number of links that depart from a book to the total number of books published afterwards. As already mentioned, this procedure was applied in our previous publication, too [21]. Figure 3 shows the normalized out-degree versus the year of publication for every book in the corpus. The normalized in-degree of each book is the percentage of the books published in the previous years that are connected to that book by at least one link.

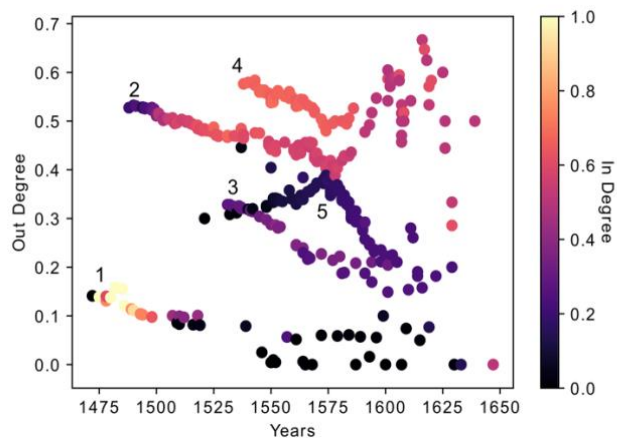

**Figure 3.** Normalized out-degree of books versus the years of their publication. The color code is based on the normalized in-degree.

There are five different branches or families of books in Fig. 3, as labeled in the plot.
- Branch 1: The first branch starts from the beginning in 1472 and has the lowest value of normalized out-degree (less than 15%) and a high value of normalized in-degree (some are around 90-100%). The low value of out-degree demonstrates the low impact of these books on books published afterwards. The first book of course has in-degree zero and the high value of in-degree of the other books in this branch implies that they are highly influenced by all the books published in earlier years.
- Branch 2: There is a jump in the normalized out-degree value to 55% that happens in 1488 and creates the second branch. The first few books in this branch have a low value of normalized in-degree but high normalized out-degree, which signals to a new type of knowledge introduced by these books. They had a high impact on future books in the corpus but were hardly influenced by previous books. A few years after the beginning of this branch, the normalized in-degree increases.
- Branches 3 and 5: The second jump happens around 1531, which is the start of two other new branches (branches 3 and 5). The value of normalized out-degree decreases from 48% to 34% and to 30% in the beginning of these branches, respectively. It is interesting that the behavior of these two branches is different. In one of them, the normalized out-degree decreases over time and in the other, it increases until the year 1574 when it shows a decreasing trend. Both branches have low values of normalized in-degree. They introduce new knowledge to the corpus and have a relatively high effect on future books, though they are less influenced by past works.
- Branch 4: The last jump occurs in 1538, the beginning of branch 4, which has the highest normalized out-degree and a comparatively high value of normalized in-degree. Books from this branch are highly influenced from past books and also strongly impact future publications.

## 2.2 Emergence of communities

The structure of the aggregated graph is shown in Fig. 4.

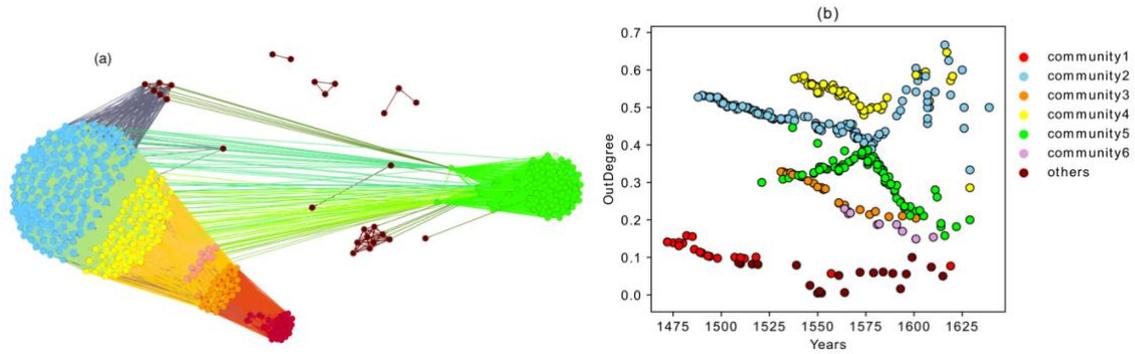

**Figure 4.** a) Aggregated graph, different communities are visualized by Louvain community detection algorithm using visualization platform Gephi [28]. b) Normalized out-degree versus years of publication colored according to the communities shown in (a). By comparing (a) and (b), it is clear that each branch in (b) is a community in aggregated graph in (a).

Six different communities are detected based on the Louvain community detection algorithm [22, 28]. Nodes in each community are highly interconnected, and the density of links between communities is smaller than the one inside communities.
The questions of interest concern the reasons behind the emergence and shaping of these communities, their common properties, the relations among them, and the role of each layer of the multiplex network in each community (Fig. 2). For this purpose, we looked first at the community structure of one of the layers and compared it with the aggregated graph. Layer se13 is the largest layer in terms of the number of nodes (321 editions), and nodes from this layer are distributed across all the communities in the aggregated network.
Layer se13 with its 28 connected components is shown in Fig. 5. By comparing the communities emerging from (i) the aggregated graph (Fig 4) and (ii) the graph from layer se13 (Fig 5), we observe that communities 1, 3, and 4 are mostly shaped by the connections present in layer se13.

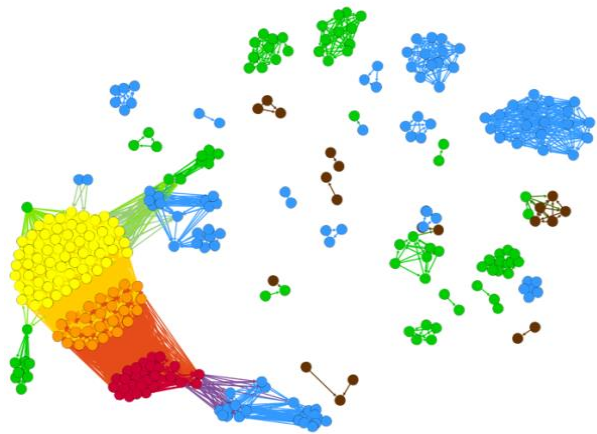

**Figure 5.** Visualization of network structure in layer se13 with 321 nodes and 28 connected components. The colors of the corresponding nodes in Fig. 5 and Fig. 4 are the same.

This observation is based on the fact that those communities emerge from the connections described in layer se13 alone, in other words, without considering the other types of semantic relations acknowledged in the remaining layers.

The connection between editions in layer se13 is based on the taxonomy same original part, which means, as mentioned, that two books are connected to each other if they contain the same original text part and the connection is directed, namely temporally oriented. The number of original parts that play a role in layer se13 is 196. Two of them are recognized as the most important ones in terms of repetition and therefore have the highest frequency. One of these (ID 100) is the main text of Johannes de Sacrobosco, *Tractatus de sphaera*. As previously mentioned, the whole corpus of historical sources is built by looking at printed books that contain that specific text part. Its relevance is therefore a consequence of the way the corpus and the dataset are built, and its high frequency does not have any general historical meaning here. The second part in terms of frequency (ID 206) is the *Propositio XXII ex libro tertio epitomae Ioannis de Regio monte in Almagestum Ptolemaei*, namely a specific demonstration (Proposition 22) from the third book of the *Abridgments* of Johannes Regiomontanus to the *Almagest* of Ptolemy. This text part is the most interesting one, as will be discussed in more details in section 3.3, too.

All the nodes from community one (red) contain part Sacrobosco's *Tractatus* (ID 100) and the nodes from community four (yellow) contain Regiomontanus' Proposition 22 (ID 206). Community three (orange), which is in between, includes nodes that have both parts. Although the nodes from community two (Fig. 4) also exist in layer se13 and are colored blue in Fig. 5 in separated components, this community is shaped mostly by the connections in layer se16. Thus, semantic relation based on the annotation of the same original part is the reason behind community two's shape.

The nodes from community five in the aggregated graph exist in layer se13 as well (which are shown in green in Fig. 5), but this community is dominated by the links coming from layer se19.

**2.3 Average age of the links to the past and future**

Now we investigate the properties of each community by calculating the average age of the links departing from each book and the average age of the links coming from sources to each book. The purpose of this study is to investigate the length of each book's influence in time. In particular, we define a method to discover a) whether each book had an immediate impact after its publication or whether a period of time had to pass until it influenced the content of other editions; b) from how far in the past a book was influenced, and c) more generally, whether any edition in the corpus refers mostly to recent publications or to the older ones. For this purpose, the average age of all the links stemming from and entering into each book are calculated and plotted versus the years of publication in Figs. 6a and 6b. The different colors in the figures mirror the communities shown in Fig. 4.

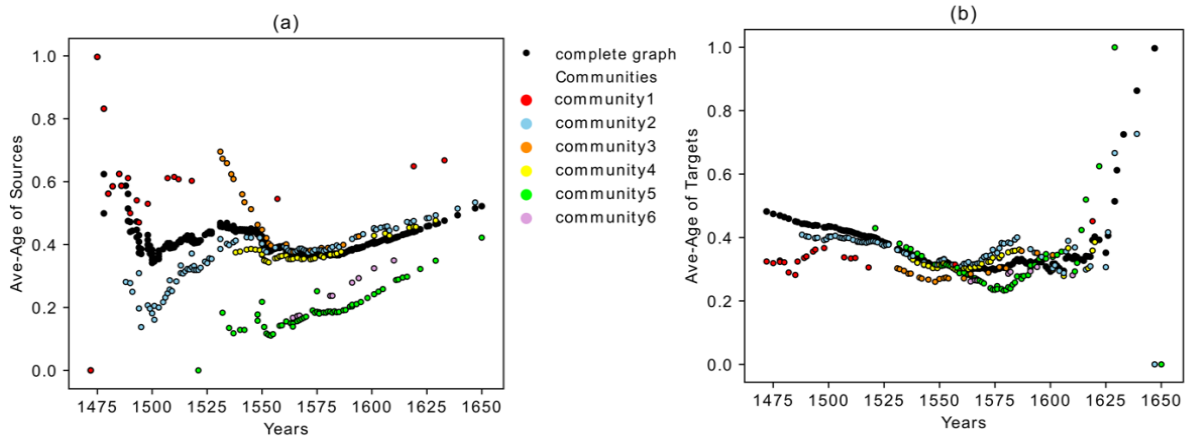

**Figure 6.** (a) Average age of the links entering into each book versus years of publication. The resulting ages of the links are normalized by dividing to the difference between the year of publication and the first year in the corpus. (b) Average age of the links that depart from each book versus the publication years. The normalization in this graph has been done by dividing average age of out-going links to the difference between last year in the corpus and the year of publication.

The black dots in Fig. 6 refer to the fully connected graph. This graph was built by connecting each book in the aggregated graph to all the books published afterwards. In this way, we made a standard reference point for our calculations. The corresponding value of each book can therefore be compared with its value in the fully connected network.

If the average age of the links which are entering into each edition is smaller than its value in the complete graph, it means that the edition has "young" references and it is mostly influenced by the editions issued shortly before its own publication. On the other hand, if the value is greater than the one of the fully connected network (or if it is above the black dots in Fig. 6a), it means that the edition refers mostly to editions older than the recent previous ones. In Fig. 6b, if the average age of the links departing from a book is smaller than its average in the complete graph, it indicates that an edition influences the editions published shortly after it.

Figure. 6a shows that community five and portions of communities two and four are below the black curve. Therefore, these books are influenced by recent publications. Contrarily, community three is above the black curve and therefore shows the deeper root of this community into the past. The books in community three are influenced by older publications. Books converging to the black curve behave like a fully connected graph and refer to books from almost all the periods in the past. Fig. 6b shows the average age of the links that depart from each book and compares it with the complete reference graph (black dots). It shows, on average, how long the effect of each book remains in the future. In this figure, communities one and three, which are under the black curve, show that the departed links from these books target into near future, meaning that their impact lasted a shorter period of time. Parts of the branches which are above the black curve, however, include the books that mostly target the far future. A peculiarity concerning these types of books is that they are not recognized in the years right after their publication; it took a certain period of time to attract attention and be referred by other younger books.

To remove the effect caused by the fact of the variable publication rate, Fig. 7 is plotted on the basis of the books' order (sorted by time). The black curve (for the fully connected reference graph) converges to the value 0.5 in both graphs and the qualitative behavior of the other curves does not change. This shows that the above-mentioned behavior of the different communities is the result of the network's topology and not of the changing publication rate.

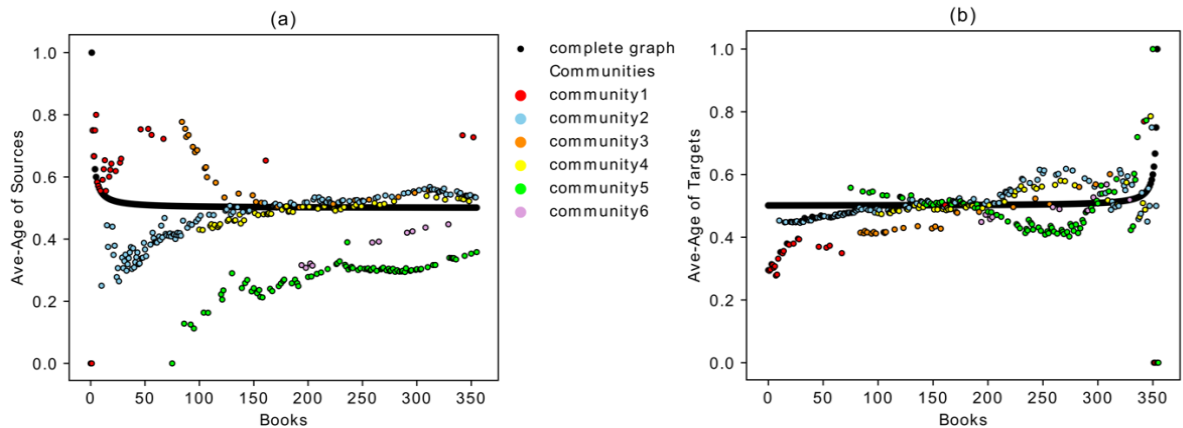

**Figure 7.** (a) Average age of the links entering into each book from past versus the temporal book order in the *Sphaera* network. (b) Average age of the links that depart from each book versus the temporal book order.

In Figs. 7a and 7b there is a convergence to the black curve between years 1549 and 1562. There are around 25 books which match with the black line with value 0.5 in both plots. We argue that these books are significant since they refer to the books from almost all periods in past, and they were also referred to by the books from almost all future periods. The distribution of the age of the links coming from past and going to the future for one of these books (ID 1903-hdl.handle.net/21.11103/sphaera.101070) in the convergence period, is depicted in Fig. 8 and compared with the corresponding distribution in the fully connected graph. As it is shown in these figures, these two distributions have a significant overlapping match with the reference graph. The same plots have been tested for all the other books in the convergence period and show similar behavior. As will be shown in the next section, the books identified in this convergence period indeed behave like a bridge between past and future. Moreover, all of them come from layer se13 (they share the same original part) and most of them also exist in layer se16 (they share the same adaption part).

Finally, we would like to highlight the following aspects concerning the different behavior of communities in terms of the average age of the sources/targets (Figs. 6 and 7).
1) Community five represents editions that refer to the most recent publications (they have younger references). Their influence to the future books, however, is high. As is shown in Figs. 7a and 7b, especially the first publications of community five have a very low value for the average age of the links *from sources* but a high value for the average age of the links *to targets*. This shows the disruptive behavior of the books of this community. They have low in-degree but high out-degree with links that go far into the future. This property is confirmed in Fig. 9 which shows the maximum age of the links from sources and to targets. As shown in Fig. 9a, the maximum age of the links from sources for all the books in community five is less than the ones for the other editions in the corpus, but the maximum age of the links to targets of these books is the highest in the corpus and shows the great influence of these books to the later publications, although they are not heavily influenced by past works. Editions from community five, therefore, introduce new knowledge into the corpus, which is highly influential during this period of study. Community five will therefore be discussed via close reading as an empirical test for our analysis in section 3.2 below.
2) Books in community three refer to the old publications, but their influence to the future does not last for a long time. This property can be easily captured in Fig. 9, which shows that the maximum age of the links from sources in these books is the highest, but the maximum age of the links to targets is very low.
3) Books in community two have two types of behaviors. The editions constituting the beginning period of the community show a disruptive behavior with low in-degree and high out-degree (Fig. 3), younger references but with a higher impact to the future (Fig. 7), and low maximum age of the links from sources but high maximum age of the links to targets (Fig. 9). Around the year 1562, this community changed its identity and was then built by editions with old references but that had a minor impact on the future publications.

4) Community one is constituted by editions published in early years covered by the corpus under examination in this paper, and their influence did not last long.
5) Books from community six were neither successful in the collection or transmission of the old tradition nor in the disruption or influence on the future books. The maximum age of the links from sources and maximum age of the links to targets of these editions are the smallest value of the corpus. They refer to very young publications and have very small impact on future books.

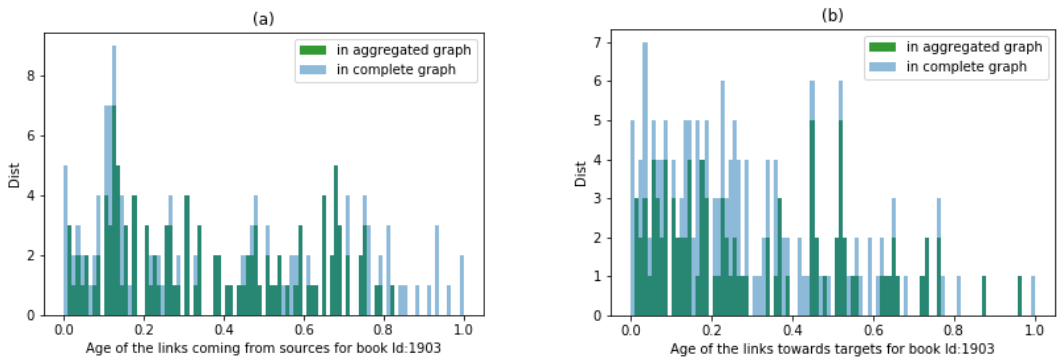

**Figure 8.** (a) Distribution of the age of the links coming from sources referring to the edition ID 1903. (b) Distribution of the age of the links to targets referring to the edition ID 1903.

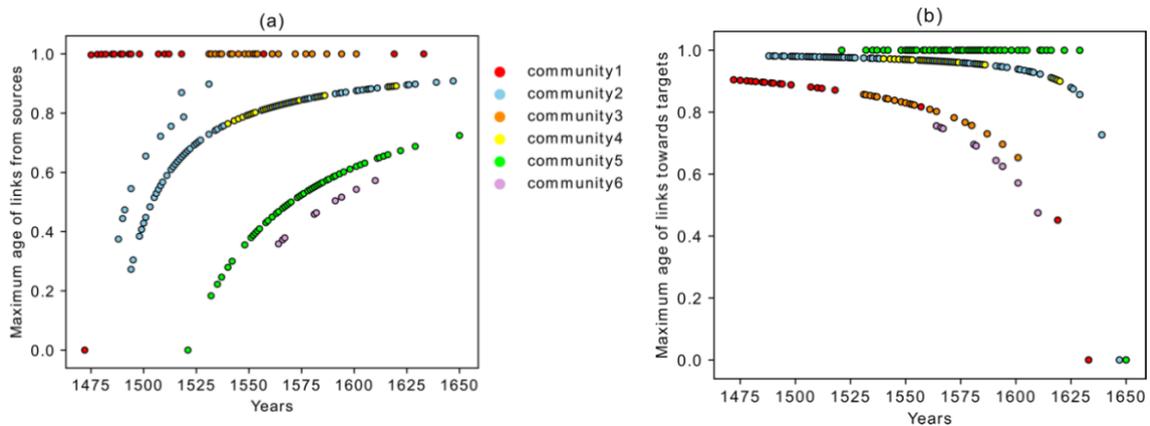

**Figure 9.** (a) Maximum age of the links from sources. (b) Maximum age of the links to targets.

## 3. Discussion

The following discussion will deepen three major historical phenomena, which the above analyses have illuminated: a) the *great transmitters*, meaning the identification of books and editions able to bridge past knowledge into their "far" future; b) the input of *disruptive innovations* and its effects; c) the fundamental mechanisms of formation of communities as exemplified by the analysis of layer se13, which connects books according to the semantic principle *same original part*.

The discussion will conclude with some remarks concerning the general process of homogenization of knowledge.

### 3.1 Great transmitters

Our analysis has shown that certain editions move very close or even overlap with the line we interpret as a reference line. We further historically interpreted this result as the identification of treatises that, more than others, transmitted knowledge over time. They collected knowledge from older works and are moreover connected with works in their distant future. As mentioned above, this phenomenon realized historically between 1549 and 1562. The analysis of the sources responsible for such a phenomenon shows that Wittenberg as a center of production remained important for a long time after its peak in 1538. Two examples will show this aspect.

The first example is concerned with a group of treatises called *Quaestiones*, a scientific genre that is here considered for the very first time in the history of transmission of knowledge. Strictly speaking they were not course textbooks but were a means for students to prepare for the exam. They were organized as questions and answers with empty pages for notes, and as such, organized the scientific knowledge in a very schematic and practical way. This form of students' book became very popular and it was particularly easy to update. In our corpus, this text tradition started in Frankfurt in 1549 with the *Quaestiones* authored by Hartmann Beyer and printed by Peter Braubach (hdl.handle.net/21.11103/sphaera.100645). After it, they were adopted by Wittenberg by 1550 through an edition of Beyer's *Questiones* printed by the Heirs of Peter I. Seitz (hdl.handle.net/21.11103/sphaera.101121), an edition that made the genre international thanks to the dominance of Wittenberg, acquired through the formation of community four as explained in [21]. As Wittenberg was observed by the other book production centers, innovations originating there were sometimes adopted by the major centers of production of books on an international level. Indeed, only one year later, in 1551, Beyer's *Quaestiones* was printed and published in Paris by the most active academic and scientific book publishers, Guillaume Cavellat (hdl.handle.net/21.11103/sphaera.101038). The broad circulation of this treatise provided the right conditions for the knowledge codified in these specific sources to remain available for a very long time. This specific edition of the *Quaestiones* was indeed republished another 17 times until 1573.

The second example primarily concerns the production of one specific treatise, also in Wittenberg by the printer Johann Krafft the Elder in 1561, namely treatise ID 1903, mentioned above (hdl.handle.net/21.11103/sphaera.101109). This work republished almost all of the text parts that were published in the 1538 edition that started the successful community four. The 1561 edition, however, shows two notable additions. First, a short work on spherical astronomy (*Themata, Quae Continent Methodicam Tractationem de Horizonte rationali ac sensibili*), originally published alone twenty years earlier, by the famous mathematical astronomer Erasmus Reinhold. In the corpus, Reinhold's text part was published 12 times, the first in 1545 (hdl.handle.net/21.11103/sphaera.100818) and the last in 1629 (hdl.handle.net/21.11103/sphaera.100303). The second is a curious and somewhat long text entitled *De ortu poetico*. It is mainly a collection of quotations concerning the times of rising and setting of the fixed stars from the classic literature (mainly ancient Latin literature, whose passages were learned by heart). The collection is finally enriched by considerations of astrological nature declared to be the original formulations of the ancient author Proclus [29]. The author of the text is unknown and is declared as such in the original early modern editions also. This text part was republished 45 times in the corpus, the first time in 1543

([hdl.handle.net/21.11103/sphaera.101029](hdl.handle.net/21.11103/sphaera.101029)) and the last in 1629 ([hdl.handle.net/21.11103/sphaera.100303](hdl.handle.net/21.11103/sphaera.100303)).

The treatise just described and that overlaps with our converging line is accompanied by three further editions whose very short descriptions will be furnished in the following section in order to have enough data to demonstrate that these treatises, seen together, indeed transferred past knowledge into the far future.

The first edition is a treatise published in the same year (1561) but in Antwerp ([hdl.handle.net/21.11103/sphaera.101058](hdl.handle.net/21.11103/sphaera.101058)). At that time, the city was on its way to becoming a very important center of book production in Europe, after Venice, Paris, and Wittenberg. From the scientific point of view, this treatise republished exactly the same 1538 Wittenberg edition, published by Joseph Klug, that gave birth to community four and mentioned already a few times. The only difference in this case is the addition of a paratext, namely a dedicatory letter of Jan Waen to Jérôme Ruffaut, abbot of St. Védaste [24]: This is a letter published in 1547 for the first time in an identical book printed by the same printer: Jean Richard. The other two texts were published in 1561 and in 1562, respectively, both in Paris by Guillaume Cavellat. These editions also republished the content of the 1538 Wittenberg edition, thereby connecting Wittenberg, Antwerp, and Paris, three major centers of production and distribution. Moreover, they also republished the collection of quotations entitled *De ortu poetico*. Finally, they added a commentary by the humanist and mathematician Élie Vinet on the treatise of Sacrobosco. This commentary was republished 23 times from 1549 ([hdl.handle.net/21.11103/sphaera.100137](hdl.handle.net/21.11103/sphaera.100137)) to 1620 ([hdl.handle.net/21.11103/sphaera.100358](hdl.handle.net/21.11103/sphaera.100358)). As shown by other research, not only did the text of Élie Vinet have considerable success over time, but he, as a scientist, was also a very central figure in the formation of scientific university textbook content [25].

To conclude the discussion concerned with the convergence period, the editions identified thanks to the above analysis conjunct trajectories of text part republications starting in 1531 in many cases and going well into the turn of the sixteenth century, almost to the end of the time interval covered by the corpus. It starts with the radical innovations of 1531 which became dominant in 1538 [21] but then it continued so that, toward the half of the century, editions that first appeared in Wittenberg were responsible for the transmission of knowledge for almost the entire century, even when reprints and republications were operated in other places and by other printers and publishers. This apparently happened thanks to two phenomena. The first is the addition of further innovations in the same circle of publishers, namely in Wittenberg. This is the case, for instance, of the text part *De ortu poetico*. The second, is the adoption from the side of Wittenberg of a new genre, the *Quaestiones*, which became an international success. The transmission of knowledge in Europe is therefore realized through the geographic, institutional, economic, intellectual and, perhaps, confessional [30] passage through Wittenberg beginning in the 1530s and lasting at least another 30 years. As a side note, it is interesting to see how the transmitting editions were accumulating text parts printed for the first time between 1531 and 1549, which means that such text parts were able to historically play the role of grand transmitter of knowledge only thanks to the fact that they were reprinted or reissued many times so that they were published during the period identified by our analysis. This consideration is relevant in view of the general tendency of historians of science to only consider the first editions of new scientific treatises relevant, that is to say, the first mention of a possible new idea. In contrast to this view, we demonstrate here that a successful scientific idea is not only relevant because of its possible inherent excellent quality but also because of its repeated appearance in a specific time and place. Therefore, further editions of the same treatises might be historically much more relevant than the first edition.

Coming finally to the discussion concerning the layers and their semantic meaning, it is interesting to notice that books on the convergence line mostly come from layers se13 and se16, meaning that great transmitters are a) original texts, in this specific case, reference texts on which other people commented, and b) commentaries related to each other because they refer to the same reference text. In conclusion, it is the mechanism that closely associates reference and commentary texts, as historically realized in sophisticated printing such as in Fig. 1, that provided the knowledge with a long temporal phase of circulation and transmission.

### 3.2. Disruptive innovations

A second important phenomenon disclosed by our analysis is related to the input of new knowledge, which has a strong impact onto successive editions. We identified the knowledge introduced by these treatises as disruptive and found their origin in community five. In this respect, the first relevant consideration is related to the layers of which community five is constituted. In particular we highlighted that all the editions of community five belong to layer 13 and layer 19 which is the new layer introduced in this research for the first time. As mentioned, these editions show their adherence to the scientific tradition of the time but also depart from it as they try to produce new content within the frames of the traditional scientific arguments. As this study therefore shows, these scientific initiatives were successful, though not immediately. The analysis of the content of the treatises that appear to have had a long temporal range of influence shows, namely, that such treatises achieved this status not at their first editions, or not only, but after a reprint or reissue.

It is interesting therefore to have a look at some of these editions, exemplarily selected, to understand which characteristics these texts had to let them have such an impact. One first aspect that seems to be relevant is that they often are adaptions (book) that, at a certain point, have been translated into a vernacular language or were originally written in the vernacular. The editions that allowed this knowledge to become disruptive, however, are not necessarily the translations into vernacular. An interesting example is the first (Latin) edition of Oronce Fine's *Cosmographia* in 1532 (hdl.handle.net/21.11103/sphaera.101190) and its second edition ten years later [31]. These editions appear to have been disruptive, and indeed starting seven further editions were issued between 1551 and 1587, three of which were in French, one in Italian, and further two in Latin.

As it is well known, writing in vernacular was a way to reach a different audience, usually more technology oriented and less connected to the classic academic milieu. This shift in the content of the treatises also seems to have been a decisive characteristic [32]. The demonstration for this is the identification of an edition as disruptive that was published in Spanish in 1535 and authored by Francisco Faleiro (hdl.handle.net/21.11103/sphaera.101182). Not only did this work contain the basic cosmological doctrine of the time but also, and more relevantly, a compact report about pressing issues of the time, such as the art of navigation and particularly the subject of the magnetic variations that made compass use on ships challenging on transoceanic travels during the sixteenth century [33].

A last example of an adaption in vernacular that turned out to be a carrier of disruptive innovation is concerned with a book by the astronomer Alessandro Piccolomini and published for the first time in 1540 in Venice directly in Italian (hdl.handle.net/21.11103/sphaera.101026). Both the first and the second edition (1548), appear to be particularly disruptive. The later seven editions (plus five editions of a revised version) include translations into French and ultimately into Latin show a dynamic that, according to our analysis, had a profound impact on the development of cosmology.

As mentioned in the analysis, community five is constituted mostly by layer se19, but not exclusively. As shown in Fig. 6, layer se13 also contributes to this community. This gives the occasion to have a closer look at the results of the analysis of layer se13 above.

## 3.3. Same original part

Layer se13 plays a fundamental role in the general network because, as it has been shown, the communities one, three, and four can all be detected in layer se13. This layer is made up of connections among editions when they contain reoccurrences of the same original text part. There are 196 original text parts that allow for the building of the graph, but especially two of them play the role of glue in these communities, as shown. Part ID 100, the *Tractatus* itself, does not generally have a historical meaning for the fact that the corpus itself has been built around this text part. However, books from community one contain only this part, as mentioned above. A very different situation is encountered when we approach part ID 206. This text part is a specific demonstration (Proposition 22) from the third book of the *Abridgments* of Johannes Regiomontanus to the *Almagest* of Ptolemy. Ptolemy was one of the greatest mathematical astronomers of antiquity and the geocentric worldview as taught at the European universities in the later medieval and early modern period was based on his magnificent (and very complex) work, the *Almagest*, as it was named in the period analyzed here. Johannes Regiomontanus was one of the most prominent fifteenth-century mathematical astronomers. In spite of the title (*Epitome-Abridgments*), his work represented a fundamental change in comparison to the *Almagest* because Regiomontanus depicted three-dimensional models of planetary orbits in order to achieve a higher precision while calculating their positions. Such models were supposed to substitute Ptolemy's geometric constructions based on eccentric and epicycles. Finally, Proposition 22 addresses the reasons for different lengths of the solar day from the temporal and spatial perspective. The subject is obviously discussed by Sacrobosco in his original treatise as well, therefore the addition of this text part—an addition decided by the publisher and printer Joseph Klug in Wittenberg [21]—has to be considered as a form of commentary and enrichment to the original text. The fact that this fragment of text was extracted from one of the greatest works on mathematical astronomy of the time was probably perceived as a major shift in the process that transformed the tradition related to Sacrobosco's treatise into a more mathematically oriented research branch. In conclusion, we can say that the formation of epistemic communities took place during an era in which the spread of knowledge related to mathematical astronomy was enriching qualitative scientific background knowledge on cosmology at the educational level.

## 3.4. Conclusive Remarks on Knowledge Economy

The historical phenomena just described can also be viewed from the perspective of knowledge economy in order to understand the process of homogenization of scientific knowledge as it was taking form during the early modern period.

First of all, we see that the most relevant text part in the formation of communities is a piece on mathematical astronomy, which indicates the relevance of the mathematization process of natural philosophy in the emergence of modern science.

Secondly, it has been shown how the most impactful innovations were those moving toward more technical aspects, which were simultaneously able to attract a larger audience of readers outside the usual institutional boundaries of the university.

These two changing aspects—math and technology—were embedded in a transmission of knowledge process that guaranteed a smooth and slow transformation along a temporal interval of over 80 years, characterized by both a continuous input of new original text parts and by a geographic (local) dominance connected to Wittenberg during the first 30 years.

Finally, regarding production of knowledge mechanisms, we have seen that the most relevant layers are se13, se16, and se19. This means that two elements, reference text and commentary on one side and adaptions (book) on the other, were the major reason for the creation of a homogenous scientific community. Therefore, the practice of commenting in and of itself was only decisive if it was in connection to specific reference texts. Furthermore, the adaptions (book) were always in reference to a specific text, as *strongly influenced by*.
We will continue the analysis of the corpus by considering other "knowledge atoms," namely scientific images and computational astronomic tables extracted from the same textbooks using machine learning techniques [34].

**Acknowledgements:** This work was partly supported by Max Planck Institute for the Physics of Complex Systems, and by the German Ministry for Education and Research as BIFOLD – Berlin Institute for the Foundations of Learning and Data (ref. 01IS18037A) and by the Max Planck Institute for the History of Science

**Contributions**

Data repository and creation: M. Vogl, F. Kräutli

Data analysis: M. Zamani, A. Tejedor

Historical interpretation: M. Valleriani

Results discussion: M. Zamani , A. Tejedor, M. Vogl, M. Valleriani, H. Kantz

Writing: M. Zamani, M. Valleriani

**Competing interests**

The authors declare no competing interests.